\documentclass[conference]{IEEEtran}
\IEEEoverridecommandlockouts
\usepackage{cite}
\usepackage{amsmath,amssymb,amsfonts}
\usepackage{algorithmic}
\usepackage{graphicx}
\usepackage{textcomp}
\usepackage{xcolor}
\usepackage{bm}
\usepackage{multirow}
\usepackage{balance}

\def\BibTeX{{\rm B\kern-.05em{\sc i\kern-.025em b}\kern-.08em
    T\kern-.1667em\lower.7ex\hbox{E}\kern-.125emX}}
\begin{document}

\title{Cross-Modal Content Inference and Feature Enrichment for Cold-Start Recommendation
\thanks{$^{*}$ indicates corresponding author.}
}

\author{
    \IEEEauthorblockN{
        Haokai Ma\textsuperscript{1}, 
        Zhuang Qi\textsuperscript{1}, 
        Xinxin Dong\textsuperscript{1}, 
        Xiangxian Li\textsuperscript{1}, 
        Yuze Zheng\textsuperscript{1}, 
        Xiangxu Meng\textsuperscript{1} 
        and Lei Meng$^*$\textsuperscript{1,2}
    }

    \IEEEauthorblockA{\textsuperscript{1} School of Software, Shandong University, Jinan, China\\}
    \IEEEauthorblockA{\textsuperscript{2} Shandong Research Institute of Industrial Technology, Jinan, China\\}

    Email: \{mahaokai, z\_qi, dongxinxin, xiangxian\_lee, zhengyuze\}@mail.sdu.edu.cn, \{mxx, lmeng\}@sdu.edu.cn
}

\maketitle

\begin{abstract}
Multimedia recommendation aims to fuse the multi-modal information of items for feature enrichment to improve the recommendation performance. However, existing methods typically introduce multi-modal information based on collaborative information to improve the overall recommendation precision, while failing to explore its cold-start recommendation performance. Meanwhile, these above methods are only applicable when such multi-modal data is available. To address this problem, this paper proposes a recommendation framework, named Cross-modal Content Inference and Feature Enrichment Recommendation (CIERec), which exploits the multi-modal information to improve its cold-start recommendation performance. Specifically, CIERec first introduces image annotation as the privileged information to help guide the mapping of unified features from the visual space to the semantic space in the training phase. And then CIERec enriches the content representation with the fusion of collaborative, visual, and cross-modal inferred representations, so as to improve its cold-start recommendation performance. Experimental results on two real-world datasets show that the content representations learned by CIERec are able to achieve superior cold-start recommendation performance over existing visually-aware recommendation algorithms. More importantly, CIERec can consistently achieve significant improvements with different conventional visually-aware backbones, which verifies its universality and effectiveness.

\end{abstract}


\section{Introduction}
\noindent
Personalized recommendation aims to capture the users' preference and provide them with the appropriate items \cite{app2, app3,app4, app5}. However, the cold-start problem is a ubiquitous challenge in recommendation, which leads to bias in the training of traditional collaborative filtering recommenders and visually-aware recommenders (e.g., Matrix Factorization (MF) \cite{MF} and Visual Bayesian Personalized Ranking (VBPR) \cite{VBPR}). That is, the randomization problem and the popularity bias problem. Content-based recommendation, which leverage the multi-modal information to improve its cold-start performance, has attracted considerable attention \cite{AdvMF, CLCRec}. However, the data in web and mobile applications is diverse and unstable \cite{app6, app1}, and the performance of existing algorithms is generally limited by the learning of heterogeneous multi-modal representations, which is not always available. Therefore, robust cross-modal inference methods for heterogeneous modal cold-start recommendation are urgently needed.

Existing cold-start recommendation methods can be categorized into two classes, based on the targets they aim at. One class is the information-based methods, which generally incorporate heterogeneous content features from the user's or item's auxiliary information \cite{socialrecommender,heterogeneousinformation,RAGAN} into the cold-start recommendation. Another class is representation-based methods, which are able to learn the fine-grained representations by dynamically optimizing the recommendation models \cite{CLCRec,MeLU,CB2CF,crossmodal}, thus improving the cold-start recommendation performance. These features are mainly obtained by encoding the items' multi-modal content information, and the accuracy and stability of these algorithms are decreased when these heterogeneous information is not available.

\begin{figure}[t]
\centering
    \includegraphics[width=0.5\textwidth]{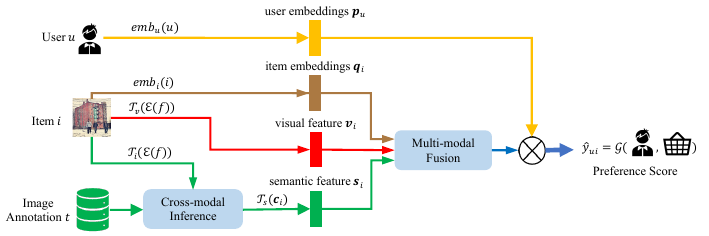}
    \caption{The main framework of CIERec. CIERec designs a novel cross-modal inference strategy to infer the cross-modal semantic content representations by introducing the privileged information into the existing visually-aware cold-start recommendation task.} 
	
    \label{fig:motivation}
    \vspace{-0.4cm}
\end{figure}

To address these issues, we present \textbf{CIERec}, a novel \underline{C}ross-modal Content \underline{I}nference and Feature \underline{E}nrichment Cold-start \underline{Rec}ommendation framework. It introduces image annotation as the privileged information to guide the mapping progress of the content representation from the visual space to the semantic space and enriches it by fusing the collaborative, visual, and inferred semantic features to improve its cold-start performance. Figure \ref{fig:motivation} shows the main framework of CIERec. Specifically, we first model the collaborative interactions in the collaborative representation learning (CRL) module. We then extract the item's uniform embedding with a traditional visual encoder(e.g. ResNet18) in the source-modal representation learning (SMRL) module. Next, to deal with the difficulty of mapping heterogeneous features, we propose a novel cross-modal inference strategy to map content representations from the visual space to the semantic space with the guidance of prior knowledge at the cross-modal representation learning (CMRL) module, which design follows a learning paradigm called learning using privileged information (LUPI) \cite{ATNet}. Finally, a multi-modal fusion method is used to integrate the user embeddings, the item embeddings, the visual features, and the inferred semantic features for the final recommendation in the multi-modal representation fusion (MRF) module. As observed, CIERec is able to alleviate the absence of heterogeneous modalities in cold-start recommendations through cross-modal inference, thereby improving the stability and accuracy of existing visually-aware cold-start recommendation models.

To validate the effectiveness of the proposed CIERec, we conduct the performance comparison on the pre-processed Allrecipes \cite{Allrecipes} and Amazon\_CDs \cite{VBPR} datasets against previous advanced visually-aware recommendation algorithms. We also conduct extensive experiments, including the ablation study to verify the effectiveness of the different components in the proposed CIERec, and the case study to visualize the distributions of heterogeneous modal representations. The experimental results demonstrate that the advantages of Tri-CDR are: (1) the privileged information can help to model the mapping relationship of content information in the visual and semantic space. (2)The dual-gating module in the CMRL module can optimize the cross-modal inference process of heterogeneous representations. (3) CIERec can further improve its cold-start performance with the fusion of multi-modal representations, allowing the downstream task to focus on the user's preference information from multiple perspectives. Furthermore, CIERec achieves consistent and significant improvements over the existing visually-aware recommendation methods with different benchmarks (e.g., MF\cite{MF} and VBPR\cite{VBPR}), proving its effectiveness and universality. Overall, the main contributions are summarized as follows:
\begin{itemize}
    \item{This paper proposes a novel Cross-modal Content Inference and Feature Enrichment Recommendation framework, CIERec, which can improve the cold-start performance of existing visually-aware recommendation methods through cross-modal semantic inference and multi-modal representation fusion.}
    \item{We design a content-enriched cross-modal inference strategy to model the heterogeneous representation inference process and extract the fine-grained multi-modal representations based on the leverage of privilege information. It can be applied as a cross-modal inference module for the general task of multi-modal representation learning.} 
    \item{We conduct extensive experiments to verify the effectiveness of the proposed CIERec on multiple datasets with different visually-aware recommendation models. The experimental results demonstrate CIERec's effectiveness, universality, and stability.}
\end{itemize}

\section{Related Work}

\subsection{Multi-modal Learning in Recommendation}
\noindent
As a common multimedia analysis method, multi-modal learning has been widely used in the fields of computer vision \cite{objectdetection, cv3, cv4,cv6,cv9}, data mining \cite{dm1,dm2,dm3,dm4}, information retrieval \cite{retrieval, ret1}, and recommendation \cite{heterogeneousinformation, PiNet, crossmodal1}. In recommendation, multimedia recommendation aims to incorporate the items' content information to model the fine-grained representation and thereby improve its performance. These methods can be classified as multi-source information embedding methods \cite{heterogeneousinformation,auxinformation,KG} and heterogeneous information inference methods \cite{PiNet,crossmodal2}, according to the available information. The multi-source information embedding methods refer to the introduction of multiple sources of heterogeneous information from the heterogeneous information networks \cite{HIN} and the knowledge graph \cite{KG} as the content information to complement the collaborative information, which can decrease the dependence of the recommendation model on the interaction information. Heterogeneous information inference methods aim to learn the mapping function from one modality space to another modality space, thus making the content representation extracted from the same item has similar distributions. These above methods typically perform cross-modal inference based on the user's interactive information \cite{crossmodal1} or multi-modal information \cite{PiNet,crossmodal2} of items, thus reducing the gap between the user's interest manifold and the visual semantic manifold. However, existing cross-modal content inference methods are rarely demonstrated against their cold-start recommendation performance, and lack the in-depth analysis of the effectiveness of the source-modal information and the inferred information for recommendation.

\subsection{Visually-aware Recommendation}
\noindent
With the development of image analysis techniques in the field of CV \cite{cv1, cv2, cv5,cv7,cv8} and NLP \cite{nlp1, nlp2, nlp3}, a series of works have validated that the incorporation of visual features of items can improve recommendation performance \cite{VBPR, Allrecipes, AdvMF, PiNet}. Inspired by the significant performance of Convolution Neural Network (CNN) in image classification, most visually-aware recommendation methods incorporate pre-extracted features as the item's embedding into the recommendation \cite{VBPR, ACF, AMR}. However, previous visually-aware recommendation methods with pre-extracted features typically utilize the items' visual features related to their category information while ignoring the users' personalized interests. Therefore, recent visually-aware recommendation methods extract visual features with an end-to-end approach to jointly optimize image encoders and recommenders. Deepstyle \cite{deepstyle} replaces collaborative representations with the real-time extraction of the items' visual features to capture the user's multi-dimensional preferences. PiNet \cite{PiNet} learns the visual features that contain both visual, semantic, and collaborative information by jointly optimizing the image encoder. However, these algorithms only take into account the visual information of the items, leading to the suboptimal performance.

\begin{figure*}[t]
    \centering
    \includegraphics[width=1\textwidth]{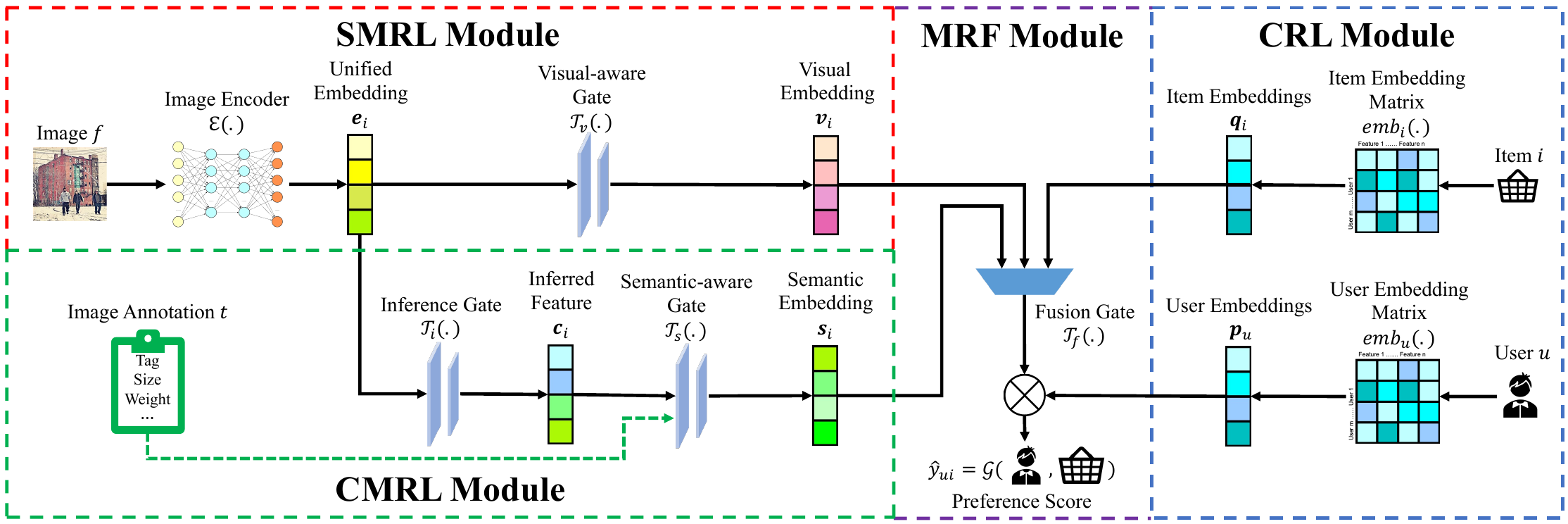}
    \caption{The overall structure of CIERec. The CRL module is used to learn the user's collaborative representation $\bm{p_u}$ and the item's collaborative representation $\bm{q_i}$; the SMRL module can be used to extract the visual feature $\bm{v_i}$; the CMRL module learns the cross-modal inferred feature $\bm{c_i}$ and then generate the semantic feature $\bm{s_i}$ from the unified feature $\bm{e_i}$; the MRF module is able to fuse the aforementioned representations and calculate the final preference score.} 
    \label{fig:overall}
    \vspace{-0.3cm}
\end{figure*}

\subsection{Cold-Start Recommendation}
\noindent
The cold start problem is common in recommendation due to the imbalance of interactions, where a few users or items dominate the interactions in the dataset. It may lead to the 'popular bias' \cite{bias}, whereby active users perform better than cold users and popular items are more likely to be recommended than cold items. Existing cold-start recommendation methods can be categorized into two categories by their targets, namely information-based methods that introduce auxiliary data (e.g., heterogeneous information \cite{heterogeneousinformation} and social information \cite{socialrecommender}) into recommendation, and the representation-based methods that modify recommendation models dynamically through meta-learning \cite{MeLU}, transfer learning \cite{crossmodal}, contrastive learning \cite{CLCRec} and other techniques. However, these algorithms have difficulty in achieving idealized recommendations performance when heterogeneous information is not available for certain modalities. To address this problem, existing studies usually deal with interaction information specifically. DropoutNet \cite{dropoutnet} adopts the dropout mechanism for the observed interactions to improve the model generalization capability. A series of studies \cite{AMF,AMR} attempt to leverage the generalization capabilities of adversarial learning, targeting at augmenting the unobserved interactions.

\section{Technique}
\subsection{Framework Overview}
CIERec introduces a cross-modal inference and feature enrichment framework to enable the enrichment of content representations through feature-level cross-modal inference. As shown in Figure \ref{fig:overall}, CIERec can be divided into four main modules, including the Collaborative Representation Learning (CRL) module, the Source-Modal Representation Learning (SMRL) module, the Cross-Modal Representation Learning (CMRL)  module, and the Multi-modal Representation Fusion (MRF) module, as illustrated in the following sections.

\subsection{Collaborative Representation Learning (CRL)}
As shown in Figure \ref{fig:overall}, CRL learns the collaborative representation $\bm{p}_u$ and the item collaborative representation $\bm{q}_i$ from the embedding matrix based on the randomly sampled user $u$ and item $i$, which is similar to the traditional collaborative filtering method. The learning process can be expressed as: 

\begin{equation}
	\begin{gathered}
		\bm{p}_u=\operatorname{emb_u} \left( u \right)\\
		\bm{q}_i=\operatorname{emb_i} \left( i \right)
	\end{gathered}
\end{equation}
where $\operatorname{emb_u}(.)$ denotes the embedding matrix of users, and $\operatorname{emb_i}(.)$ denotes the embedding matrix of items.

\vspace{-0.1cm}
\subsection{Source-Modal Representation Learning (SMRL)}
CIERec generates visual feature $ \bm{v}_i$ in the SMRL module to complement the collaborative representation $\bm{q}_i$. Specifically, as illustrated in Figure \ref{fig:overall}, the SMRL module extracts the uniform embedding $\mathcal{E}(f) \mapsto \bm{e}_{i}$ from the image $f$ through the visual encoder $\mathcal{E}(.)$. Inspired by the dual-gating mechanism \cite{PiNet}, we develop a novel multimodal-gating function to generate the visual feature $\mathcal{T}_v\left(\bm{e}_i\right) \mapsto \bm{v}_i$ by mapping with the visual-aware gate $\mathcal{T}_v$, and constrains its optimization with the gradient-regularization gate $\mathcal{R}$. The overall computational equation can be expressed as:
\begin{equation}
	\bm{v}_i=\mathcal{T}_v\left(\mathcal{E}\left( f \right)\right)
\end{equation}

There is significant heterogeneity between the image regions focused by visual feature $\bm{v}_i$ and semantic feature $\bm{s}_i$, which is difficult to be directly mapped from the uniform embedding $\bm{e}_i$. Therefore, CIERec proposes a visual-aware gate to control the delivery of visual information. The visual-aware gate $\mathcal{T}_v(.)$ of CIERec introduces a self-learning gate vector $\bm{g}_v$ and the user representation $\bm{p}_u$ in addition to the uniform embedding $\bm{e}_i$, and maps them to the visual feature space, which is defined as follows:
\begin{equation}
	\bm{v}_i\!=\!\mathcal{T}_v\left(\bm{e}_{i}\right)\!=\!\operatorname{MLP}^v\left(\frac{\bm{e}_i \odot \delta\left(\bm{p}_{u} \| \bm{e}_{i} \| \bm{g}_{v}\right)}{\operatorname{max}(\left\|\bm{e}_i \odot \delta\left(\bm{p}_{u} \| \bm{e}_{i} \| \bm{g}_{v}\right)\right\|_{2}, \epsilon)}\right)
\end{equation}
where $\|$ denotes the concatenate operator, $\odot$ denotes the dot product operator, $\left\|.\right\|_{2}$ denotes the $\ell_2$ regularization function, $\epsilon = 1e^{-12}$ denotes a small value to avoid division by zero, $\delta\left(.\right)$ denotes a fully connected layer and $\operatorname{MLP}^v(.)$ denotes a fully connected layer followed by the $\operatorname{LeakyReLU}$ activation function. We further describe the gradient-regularization gate $\mathcal{R}$ in Sec. \ref{sec:gradient gate}.

\subsection{Cross-modal Representation Learning (CMRL)}
In addition to generating the visual feature $\bm{v}_i$, we firstly performs cross-modal semantic inference based on the inference gate $\mathcal{T}_i$ to learn the inference feature $\bm{c}_i$ from the unified embedding $\bm{e}_i$; And then we conducts semantic fusion based on the prior knowledge, that is, fuses the semantic knowledge $s$ of image annotation $t$ with the cross-modal inference feature $\bm{c}_i$ to generate semantic feature $\bm{s}_i$ via the semantic-aware gate $\mathcal{T}_s$, which in turn complements the multimodal representation.

\subsubsection{Semantic Inference}
In preliminary experiments, we found that the incorporation of semantic information can significantly improve the recommendation performance of traditional visually-aware recommendation methods. Existing multimedia recommendation methods generally rely on the modal richness of the recommendation datasets, which performs poorly when semantic information is unavailable. To address this issue, the CMRL module generates the cross-modal inference feature $\mathcal{T}_i\left(\bm{e}_i\right)\mapsto \bm{c}_i$ by mapping from the unified embedding $\bm{e}_i$ with the inference gate $\mathcal{T}_i(.)$ and uniformly optimizes it with the gradient-regularization gate $\mathcal{R}$. It is defined as follows:
\begin{equation}
	\bm{c}_{i}=\frac{\bm{e}_i \odot \delta\left(\bm{e}_i \| \bm{g}_s\right)}{\operatorname{max}(\left\|\bm{e}_i \odot \delta\left(\bm{e}_i \| \bm{g}_s\right)\right\|_{2}, \epsilon)}
\end{equation}
where $\bm{g}_s$ denotes the learnable inference gate vector, $\|$ denotes the concatenate operator, $\left\|.\right\|_{2}$ denotes the $\ell_2$ regularization function, $\epsilon = 1e^{-12}$ denotes a small value to avoid division by zero and $\bm{c}_i$ denotes the inferred feature.

\subsubsection{Semantic fusion for privileged information}
The inferred feature $\bm{c}_i$ derived from the uniform embedding $\bm{e}_i$ contains a certain amount of redundant information, leading to bias in the subsequent recommendation process. To address this issue, the CMRL module also introduces the prior semantic information to improve the model's ability to represent semantic modalities. The CMRL module uses LSTM \cite{lstm} as the building block for the semantic element encoder to capture the relationship between the prior semantic representation and the inferred representation. The process of extracting the semantic embedding $\mathcal{T}_s\left(\bm{c}_i\right)\mapsto \bm{s}_i$ can be represented as:

\begin{equation}
	\bm{s}_i = \operatorname{LSTM}(\bm{c}_i \| \hat{\bm{s}}_{i})
\end{equation}
where $\hat{\bm{s}}_i$ denotes the privileged knowledge encoded from the image annotation, $\operatorname{LSTM}\left(.\right)$ denotes the the LSTM operator, $\bm{s}_i$ denotes the calculated semantic embedding.

\subsubsection{Gradient-regularization Gate}
\label{sec:gradient gate}
CIERec fuses the gradients of the two heterogeneous representations with the gradient-regularization gate $\mathcal{R}$, aiming to allow the visual encoder to trade off the visual feature $\bm{v}_i$ and the inferred semantic feature $\bm{s}_i$ from the unified embedding $\bm{e}_i$ during the backpropagation. The visual gradient-regularization gate is implemented based on the deep Q-network (DQN) \cite{mnih2015human} approach. It selects the action $\bm{s}^{\left( t\right)}$ via a classifier $\operatorname{MLP}_{DQN}^v$ to map the 5D state vector of $\bm{s}^{\left( t-1\right)}$ at batch $t$, where $\operatorname{MLP}_{DQN}^v$ denotes a fully-connected layer followed by a Softmax activation function. And then it penalizes this action by the $J_{v}^{(t)}=\exp \left(-\mathcal{L}_{v}^{(t+1)}\right)$ obtained by the feedback from the recommendation model, with the loss function as following:
\begin{equation}
	\mathcal{L}_{dqn}=-\log \sigma\left(\bm{s}_{\max }^{(t)} \cdot J_{v}^{(t)}\right)
\end{equation}
where $\!\bm{s}_{\max}^{(t)}\!$ is the probablity of selecting $\!\bm{s}^{(t)}\!$, and $\!\sigma\left(.\right)\!$ is the $\operatorname{Sigmoid}$ Function. Similarly, we have $\!J_{s}^{(t)}\!=\!\exp\! \left(\!-\mathcal{L}_{s}^{(t+1)}\!\right)$ for the semantic gradient-regularization gate.


\subsection{Multimodal Representation Fusion (MRF) Module}
To demonstrate the universality of CIERec, we take the traditional collaborative filtering algorithms MF and VBPR as the backbone models for this module, which represent user and product as an embedding vector, with the core idea of estimating the user's preference as the inner product of their embedding vectors\cite{MF}. In addition to the user representation $\bm{p}_u$ and the collaborative representation $\bm{q}_i$, MRF module also receives the visual embedding $\bm{v}_i$ and the semantic embedding $\bm{s}_i$ for recommendation. The fusion operation of multi-modal representations can be represented as:
\begin{equation}
	\bm{f}_i=\mathcal{T}_f\left(\bm{q}_{i}, \bm{v}_i, \bm{s}_i\right)=\operatorname{MLP}\left(\bm{q}_i \| \bm{v}_i \| \bm{s}_i\right)
\end{equation}
where $\mathcal{T}_f$ denotes the fusion gate of the multi-modal representations, $\operatorname{MLP}\left(.\right)$ denotes a fully connected layer with a LeakyReLU activation function, and $\bm{f}_i$denotes the multi-modal fused representation of item $i$. 
The process of calculating preference scores for the BPR-MF and VBPR algorithms is defined as following:
\begin{equation}
	\begin{gathered}
		\hat{y}_{MF}=\alpha+\beta_{u}+\beta_{i}+\bm{p}_u^{\top} \bm{f}_{i} \\
		\hat{y}_{VBPR}=\alpha+\beta_{u}+\beta_{i}+\beta_{c}+\bm{p}_u^{\top} \bm{f}_{i}+\bm{a}_u^{\top} \bm{c}_{i}
	\end{gathered}
\end{equation}
where $\alpha$ denotes the self-learning global bias, $\beta_u$, $\beta_i$ and $\beta_c$ denote the self-learning bias of user $u$, item $i$ and the content $c$ respectively, $\bm{a}_u$ denotes the implicit representation, ${\hat{y}}_{MF}$ and ${\hat{y}}_{VBPR}$ denote the preference score of MF and VBPR.

\section{Experiments}

\subsection{Experimental Setup}
\subsubsection{Datasets}
\begin{table}[t]
    \centering
    \caption{Statistics of the experimented datasets.}
    \label{table:dataset}
    \renewcommand\arraystretch{1.5}  
    \resizebox{\linewidth}{!}{
        \begin{tabular}{|c|c|c|c|c|c|}
            \hline
            Datasets & Users & Items & Interactions & Elements & Sparsity \\ \hline
            Allrecipes & 68,768 & 45,630 & 1,093,845 & 2,736             & 99.97\%   \\ \hline
            Amazon\_CDs & 67,282 & 40,314 & 752,724 & 467 & 99.97\%   \\ \hline
    \end{tabular}}
	
\end{table}

\begin{table*}[t] 
    \centering \caption{Performance comparison of CIERec with existing baseline algorithms on Amazon\_CDs and Allrecipes datasets.}
    \label{tab:performance}
    \renewcommand\arraystretch{1.5}  
    \resizebox{\linewidth}{!}{
		\begin{tabular}{|c|c|cccccc|cccccc|}
			\hline
			\multirow{3}{*}{Methods} &
			\multirow{3}{*}{Algorithms} &
			\multicolumn{6}{c|}{Amazon\_CDs Dataset} &
			\multicolumn{6}{c|}{Allrecipes Dataset} \\ \cline{3-14} 
			&
			&
			\multicolumn{2}{c|}{Cold} &
			\multicolumn{2}{c|}{Warm} &
			\multicolumn{2}{c|}{All} &
			\multicolumn{2}{c|}{Cold} &
			\multicolumn{2}{c|}{Warm} &
			\multicolumn{2}{c|}{All} \\ \cline{3-14} 
			&
			&
			\multicolumn{1}{c|}{R@10} &
			\multicolumn{1}{c|}{NDCG@10} &
			\multicolumn{1}{c|}{R@10} &
			\multicolumn{1}{c|}{NDCG@10} &
			\multicolumn{1}{c|}{R@10} &
			NDCG@10 &
			\multicolumn{1}{c|}{R@10} &
			\multicolumn{1}{c|}{NDCG@10} &
			\multicolumn{1}{c|}{R@10} &
			\multicolumn{1}{c|}{NDCG@10} &
			\multicolumn{1}{c|}{R@10} &
			NDCG@10 \\ \hline
			\multirow{2}{*}{Backbone} &
			MF &
			\multicolumn{1}{c|}{0.1719} &
			\multicolumn{1}{c|}{0.1483} &
			\multicolumn{1}{c|}{0.1763} &
			\multicolumn{1}{c|}{0.3678} &
			\multicolumn{1}{c|}{0.1734} &
			0.2233 &
			\multicolumn{1}{c|}{0.2427} &
			\multicolumn{1}{c|}{0.1936} &
			\multicolumn{1}{c|}{0.1962} &
			\multicolumn{1}{c|}{0.4555} &
			\multicolumn{1}{c|}{0.2275} &
			0.2793 \\ \cline{2-14} 
			&
			VBPR &
			\multicolumn{1}{c|}{0.1724} &
			\multicolumn{1}{c|}{0.1533} &
			\multicolumn{1}{c|}{0.1933} &
			\multicolumn{1}{c|}{0.4087} &
			\multicolumn{1}{c|}{0.1796} &
			0.2411 &
			\multicolumn{1}{c|}{0.2539} &
			\multicolumn{1}{c|}{0.2050} &
			\multicolumn{1}{c|}{0.2204} &
			\multicolumn{1}{c|}{0.4948} &
			\multicolumn{1}{c|}{0.2429} &
			0.2999 \\ \hline
			\multirow{5}{*}{\begin{tabular}[c]{@{}c@{}}Multi-modal\\ Learning\end{tabular}} &
			MF(Image) &
			\multicolumn{1}{c|}{0.1634} &
			\multicolumn{1}{c|}{0.1459} &
			\multicolumn{1}{c|}{0.1930} &
			\multicolumn{1}{c|}{0.3994} &
			\multicolumn{1}{c|}{0.1736} &
			0.2331 &
			\multicolumn{1}{c|}{0.2450} &
			\multicolumn{1}{c|}{0.1975} &
			\multicolumn{1}{c|}{0.2085} &
			\multicolumn{1}{c|}{0.4782} &
			\multicolumn{1}{c|}{0.2330} &
			0.2894 \\ \cline{2-14} 
			&
			MF(Semantics) &
			\multicolumn{1}{c|}{0.1843} &
			\multicolumn{1}{c|}{0.1589} &
			\multicolumn{1}{c|}{0.2107} &
			\multicolumn{1}{c|}{0.4086} &
			\multicolumn{1}{c|}{0.1933} &
			0.2447 &
			\multicolumn{1}{c|}{0.2376} &
			\multicolumn{1}{c|}{0.1886} &
			\multicolumn{1}{c|}{0.2103} &
			\multicolumn{1}{c|}{0.4683} &
			\multicolumn{1}{c|}{0.2286} &
			0.2801 \\ \cline{2-14} 
			&
			VECF &
			\multicolumn{1}{c|}{0.1796} &
			\multicolumn{1}{c|}{0.1555} &
			\multicolumn{1}{c|}{0.1989} &
			\multicolumn{1}{c|}{0.4062} &
			\multicolumn{1}{c|}{0.1871} &
			0.2426 &
			\multicolumn{1}{c|}{0.2608} &
			\multicolumn{1}{c|}{0.2062} &
			\multicolumn{1}{c|}{0.2284} &
			\multicolumn{1}{c|}{0.4977} &
			\multicolumn{1}{c|}{0.2502} &
			0.3016 \\ \cline{2-14} 
			&
			HAFR-non-i &
			\multicolumn{1}{c|}{0.1872} &
			\multicolumn{1}{c|}{0.1619} &
			\multicolumn{1}{c|}{0.2052} &
			\multicolumn{1}{c|}{0.4174} &
			\multicolumn{1}{c|}{0.1934} &
			0.2498 &
			\multicolumn{1}{c|}{0.2600} &
			\multicolumn{1}{c|}{0.2095} &
			\multicolumn{1}{c|}{0.2290} &
			\multicolumn{1}{c|}{0.5064} &
			\multicolumn{1}{c|}{0.2499} &
			0.3067 \\ \cline{2-14} 
			&
			PiNet &
			\multicolumn{1}{c|}{0.2100} &
			\multicolumn{1}{c|}{0.1833} &
			\multicolumn{1}{c|}{0.2240} &
			\multicolumn{1}{c|}{0.4367} &
			\multicolumn{1}{c|}{0.2148} &
			0.2705 &
			\multicolumn{1}{c|}{0.2770} &
			\multicolumn{1}{c|}{0.2209} &
			\multicolumn{1}{c|}{0.2377} &
			\multicolumn{1}{c|}{0.5106} &
			\multicolumn{1}{c|}{0.2641} &
			0.3158 \\ \hline
			\multirow{4}{*}{\begin{tabular}[c]{@{}c@{}}Cold-Start\\ Learning\end{tabular}} &
			DropoutNet &
			\multicolumn{1}{c|}{0.1744} &
			\multicolumn{1}{c|}{0.1507} &
			\multicolumn{1}{c|}{0.1876} &
			\multicolumn{1}{c|}{0.3790} &
			\multicolumn{1}{c|}{0.1789} &
			0.2292 &
			\multicolumn{1}{c|}{0.2503} &
			\multicolumn{1}{c|}{0.2002} &
			\multicolumn{1}{c|}{0.2043} &
			\multicolumn{1}{c|}{0.4619} &
			\multicolumn{1}{c|}{0.2353} &
			0.2859 \\ \cline{2-14} 
			&
			AMF &
			\multicolumn{1}{c|}{0.1717} &
			\multicolumn{1}{c|}{0.1484} &
			\multicolumn{1}{c|}{0.1818} &
			\multicolumn{1}{c|}{0.3750} &
			\multicolumn{1}{c|}{0.1752} &
			0.2263 &
			\multicolumn{1}{c|}{0.2494} &
			\multicolumn{1}{c|}{0.2006} &
			\multicolumn{1}{c|}{0.2069} &
			\multicolumn{1}{c|}{0.4734} &
			\multicolumn{1}{c|}{0.2355} &
			0.2899 \\ \cline{2-14} 
			&
			AMR &
			\multicolumn{1}{c|}{0.1826} &
			\multicolumn{1}{c|}{0.1620} &
			\multicolumn{1}{c|}{0.2000} &
			\multicolumn{1}{c|}{0.4160} &
			\multicolumn{1}{c|}{0.1886} &
			0.2494 &
			\multicolumn{1}{c|}{0.2619} &
			\multicolumn{1}{c|}{0.2106} &
			\multicolumn{1}{c|}{0.2238} &
			\multicolumn{1}{c|}{0.4977} &
			\multicolumn{1}{c|}{0.2494} &
			0.3046 \\ \cline{2-14} 
			&
			CLCRec &
			\multicolumn{1}{c|}{0.1963} &
			\multicolumn{1}{c|}{0.1723} &
			\multicolumn{1}{c|}{0.2087} &
			\multicolumn{1}{c|}{0.4391} &
			\multicolumn{1}{c|}{0.2005} &
			0.2641 &
			\multicolumn{1}{c|}{0.2571} &
			\multicolumn{1}{c|}{0.2056} &
			\multicolumn{1}{c|}{0.2327} &
			\multicolumn{1}{c|}{0.5012} &
			\multicolumn{1}{c|}{0.2491} &
			0.3024 \\ \hline
			\multirow{2}{*}{\begin{tabular}[c]{@{}c@{}}Cross-modal\\ Learning\end{tabular}} &
			CIERec(MF) &
			\multicolumn{1}{c|}{\underline{0.2257}} &
			\multicolumn{1}{c|}{\underline{0.1981}} &
			\multicolumn{1}{c|}{\underline{0.2461}} &
			\multicolumn{1}{c|}{\underline{0.4706}} &
			\multicolumn{1}{c|}{\underline{0.2327}} &
			\underline{0.2918} &
			\multicolumn{1}{c|}{\underline{0.2789}} &
			\multicolumn{1}{c|}{\underline{0.2228}} &
			\multicolumn{1}{c|}{\underline{0.2427}} &
			\multicolumn{1}{c|}{\underline{0.5165}} &
			\multicolumn{1}{c|}{\underline{0.2670}} &
			\underline{0.3190} \\ \cline{2-14} 
			&
			CIERec(VBPR) &
			\multicolumn{1}{c|}{\textbf{0.2376}} &
			\multicolumn{1}{c|}{\textbf{0.2091}} &
			\multicolumn{1}{c|}{\textbf{0.2602}} &
			\multicolumn{1}{c|}{\textbf{0.4957}} &
			\multicolumn{1}{c|}{\textbf{0.2454}} &
			\textbf{0.3076} &
			\multicolumn{1}{c|}{\textbf{0.2832}} &
			\multicolumn{1}{c|}{\textbf{0.2269}} &
			\multicolumn{1}{c|}{\textbf{0.2447}} &
			\multicolumn{1}{c|}{\textbf{0.5209}} &
			\multicolumn{1}{c|}{\textbf{0.2706}} &
			\textbf{0.3232} \\ \hline
	\end{tabular}}
\end{table*}

We conduct experiments on two real-world recommendation datasets, where Allrecipes was constructed by Gao \cite{Allrecipes} and Amazon\_CDs was extracted from the original Amazon dataset \cite{VBPR} to meet the needs for this task. To verify the effectiveness of CIERec in alleviating the cold-start problems in recommendation, we divided Allrecipes and Amazon\_CDs into two parts with the interaction boundary of 3, i.e., the cold-start set with few interactions and the warm-start set with a relatively large number of interactions. Table \ref{table:dataset} summarizes the statistics of the datasets. Both datasets follow the data partitioning method used in Allrecipes, where the train set includes the earliest 60\% of the interaction data for each user, the test set includes the latest 30\% of the interaction data, and the remaining 10\% is used as the valid set.

\subsubsection{Evaluation measures}
Following the classical cold-start recommendation works \cite{CLCRec, metric2}, two widely-used metrics are adopted to evaluate the performance of the cold start recommendation, including Recall (R) and Normalized discounted cumulative gain (NDCG)\cite{CLCRec}. Following \cite{PiNet}, we randomly select one negative sample for each positive sample in training, while in testing five hundred items are randomly selected as the negative samples (have no interaction with the user) from the dataset along with all positive items (have interactions with the user) to form the ranking candidate for each user. R@K and NDCG@K calculate the performance of positive samples in the Top-k ranking items for all sampled items. To alleviate the problem of randomness, we repeat the evaluation process five times and report the average value.

\subsubsection{Implementation details}
Based on the efficiency and performance of ResNet18 in recommendation  \cite{ma2021comparative} and prediction \cite{li2020wavelet}, CIERec used it as the visual encoder to extract the uniform representations with the dimension of 512. The cross-modal recommendation model was optimized by Adagrad with the learning rate from 0.0001 to 0.5, and the DQN model was optimized by Adam with the learning rate from 0.00001 to 0.005. Both the batch size and the dimension of the CIERec were selected in {32, 64, 28, 256}, and these optimizers are decayed proportionally for every four epochs, with the decay rate chosen from {0.1, 0.5}.

\subsubsection{Baselines}
\label{sec:baseline}
We compare CIERec with both multi-modal learning models and cold-start learning models:

\begin{itemize}
  \item MF \cite{MF} is a classical recommendation model to utilize the implicit feedback information with Bayesian Personalized Ranking.
  \item VBPR \cite{VBPR} is a factorization model to incorporate pre-extracted visual features into recommendation.
  \item MF(Image/Semantic) \cite{MF} is a variant of MF where we replace the collaborative embedding with visual or semantic feature. For fair comparisons, we use the same pre-extracted visual features and semantic feature for training, noted as MF(Image) and MF(Semantic).
  \item VECF \cite{VECF} capture the visual and textual features by a multi-modal attention network seamlessly.
  \item HAFR-non-i \cite{Allrecipes} learns the user’s preference via the visual images, ingredients and the collaborative information of the interacted recipes.
  \item PiNet \cite{PiNet} is a heterogeneous multi-task learning framework that learns visual features containing with semantic and collaborative information, which is also the baseline method for the proposed CIERec.
  \item DropoutNet \cite{dropoutnet} learn a DNN-based latent model via the dropout mechanism based on the idea that cold start is equivalent to the missing data problem.
  \item AMF \cite{AMF} learns the effective collaborative feature via adding gradient-based perturbations to item embedding.
  \item AMR \cite{AMR} adds random-based perturbations and gradient-based perturbations to the pre-extracted visual features to model the effective visual information.
  \item CLCRec \cite{CLCRec} is a SOTA model in cold-start recommendation that optimizes the dependence between items' embedding and content information via contrastive learning.
\end{itemize}

\begin{table*}[t]
    \centering \caption{Recall@10 on ablation study of CIERec(MF) and CIERec(VBPR).}
    \label{tab:ablation}
     \renewcommand\arraystretch{1.5}  
     \resizebox{\linewidth}{!}{
		\begin{tabular}{|c|cccccc|cccccc|}
			\hline
			\multirow{3}{*}{Algorithms} &
			\multicolumn{6}{c|}{CIERec(MF)} &
			\multicolumn{6}{c|}{CIERec(VBPR)} \\ \cline{2-13} 
			&
			\multicolumn{3}{c|}{Amazon\_CDs} &
			\multicolumn{3}{c|}{Allrecipes} &
			\multicolumn{3}{c|}{Amazon\_CDs} &
			\multicolumn{3}{c|}{Allrecipes} \\ \cline{2-13} 
			&
			\multicolumn{1}{c|}{Cold} &
			\multicolumn{1}{c|}{Warm} &
			\multicolumn{1}{c|}{All} &
			\multicolumn{1}{c|}{Cold} &
			\multicolumn{1}{c|}{Warm} &
			All &
			\multicolumn{1}{c|}{Cold} &
			\multicolumn{1}{c|}{Warm} &
			\multicolumn{1}{c|}{All} &
			\multicolumn{1}{c|}{Cold} &
			\multicolumn{1}{c|}{Warm} &
			All \\ \hline
			Base &
			\multicolumn{1}{c|}{0.1634} &
			\multicolumn{1}{c|}{0.1930} &
			\multicolumn{1}{c|}{0.1736} &
			\multicolumn{1}{c|}{0.2450} &
			\multicolumn{1}{c|}{0.2085} &
			0.2330 &
			\multicolumn{1}{c|}{0.1724} &
			\multicolumn{1}{c|}{0.1933} &
			\multicolumn{1}{c|}{0.1796} &
			\multicolumn{1}{c|}{0.2539} &
			\multicolumn{1}{c|}{0.2204} &
			0.2429 \\ \hline
			Base+CI &
			\multicolumn{1}{c|}{0.1688} &
			\multicolumn{1}{c|}{0.1766} &
			\multicolumn{1}{c|}{0.1715} &
			\multicolumn{1}{c|}{0.2405} &
			\multicolumn{1}{c|}{0.2071} &
			0.2296 &
			\multicolumn{1}{c|}{0.1685} &
			\multicolumn{1}{c|}{0.1850} &
			\multicolumn{1}{c|}{0.1742} &
			\multicolumn{1}{c|}{0.2445} &
			\multicolumn{1}{c|}{0.2130} &
			0.2342 \\ \hline
			Base+CI+TA &
			\multicolumn{1}{c|}{0.1906} &
			\multicolumn{1}{c|}{0.2043} &
			\multicolumn{1}{c|}{0.1953} &
			\multicolumn{1}{c|}{0.2536} &
			\multicolumn{1}{c|}{0.2177} &
			0.2419 &
			\multicolumn{1}{c|}{0.1843} &
			\multicolumn{1}{c|}{0.2135} &
			\multicolumn{1}{c|}{0.1943} &
			\multicolumn{1}{c|}{0.2739} &
			\multicolumn{1}{c|}{0.2368} &
			0.2617 \\ \hline
			Base+CI+TA+GR &
			\multicolumn{1}{c|}{0.2006} &
			\multicolumn{1}{c|}{0.2145} &
			\multicolumn{1}{c|}{0.2054} &
			\multicolumn{1}{c|}{0.2718} &
			\multicolumn{1}{c|}{0.2325} &
			0.2589 &
			\multicolumn{1}{c|}{0.2059} &
			\multicolumn{1}{c|}{0.2184} &
			\multicolumn{1}{c|}{0.2102} &
			\multicolumn{1}{c|}{0.2777} &
			\multicolumn{1}{c|}{0.2407} &
			0.2656 \\ \hline
			Base+CI+TA+GR+PI &
			\multicolumn{1}{c|}{\textbf{0.2257}} &
			\multicolumn{1}{c|}{\textbf{0.2461}} &
			\multicolumn{1}{c|}{\textbf{0.2327}} &
			\multicolumn{1}{c|}{\textbf{0.2789}} &
			\multicolumn{1}{c|}{\textbf{0.2427}} &
			\textbf{0.2670} &
			\multicolumn{1}{c|}{\textbf{0.2376}} &
			\multicolumn{1}{c|}{\textbf{0.2602}} &
			\multicolumn{1}{c|}{\textbf{0.2454}} &
			\multicolumn{1}{c|}{\textbf{0.2832}} &
			\multicolumn{1}{c|}{\textbf{0.2447}} &
			\textbf{0.2706} \\ \hline
	\end{tabular}}
\end{table*}

\begin{figure*}[t]
    \centering
    \includegraphics[width=1\textwidth]{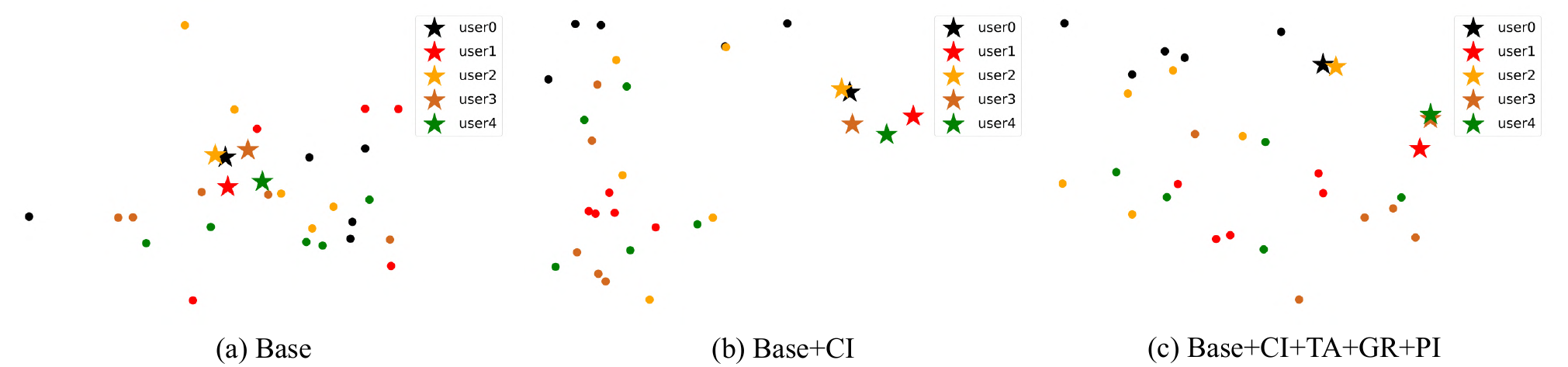}
    \caption{Visualization of the learned t-SNE transformed representations derived from (a) Base, (b) Base+CI, and (c) Base+CI+TA+GR+PI mentioned in Sec.\ref{sec:ablation}. Each star denotes a user and the points with the same color indicate the interacted item.} 
	
    \label{fig:tsne}
\end{figure*}

\subsection{Performance Comparison}
In this section, we compare CIERec with the algorithm mentioned in section \ref{sec:baseline} for performance comparison. For each algorithm, we have fine-tuned its hyper-parameters to obtain its best performance in the experiments. It can be observed from Table \ref{tab:performance} that:

\begin{itemize}
    \item{\verb||} MF \cite{MF} with content features achieves consistent improvements over MF on most datasets and metrics, whereas replacing the collaborative information only with content information from a single modality can lead to a decrease in cold-start performance (MF(Image) achieved a 4.9\% decrease on Amazon\_CDs and MF(Semantics) obtained a 1.9\% decrease on Allrecipes), demonstrating the necessity of fine-grained processing of multi-modal information in cold-start recommendations. 
    \item{\verb||} PiNet \cite{PiNet} outperforms the traditional visually-aware recommendation methods (e.g., VECF \cite{VECF} and HAFR-non-i \cite{Allrecipes}). This is mainly owing to the fact that it introduces semantic information in addition to visual information to regularize the learning process of the content representation.
    \item{\verb||} CIERec outperforms existing multi-modal learning methods in all performance metrics, which validates that CIERec can bring significant and consistent improvement over existing recommendation methods by fusing the user representation, the collaborative representation, and the multi-modal content information obtained from cross-modal inference.
    \item{\verb||} CIERec achieves higher cold-start improvements on the Amazon\_CDs, which is due to its fewer classes of semantic elements. That is, the same semantic element category corresponds to more items, thus making the cross-modal inferred semantic information more representative. It also proves the importance of semantic information in cold-start recommendation.
\end{itemize}

\subsection{Ablation Study}
\label{sec:ablation}
In addition to the overall performance comparison, we further explore the effectiveness of the combinations of components in CIERec. Specifically, we use Base, CI, TA, GR and PI to represent using the components of plain recommendation model, cross-modal inference, task aware gate, gradient regularization and privileged information respectively. The ablation results are shown in Table \ref{tab:ablation} and we have the following findings:

\begin{itemize}
    \item{\verb||} For 'CI', that is, cross-modal inference without any constraints, the application of naive inference in recommendations leads to a significant degradation of its performance. This may explained by the fact that a large number of visual information is lost in the cross-modal transformation.
    \item{\verb||} For 'TA' and 'GR', 'TA' facilitates the modeling of mapping relationships between heterogeneous modalities through task-aware gates in each modality; while 'GR' can help to learn the optimization direction of heterogeneous representations through gradient-regularization gates within the constraints of reinforcement learning. CIERec can further improve its performance by combining these components.
    \item{\verb||} The 'PI' component is able to introduce the prior knowledge (e.g., the textual annotations of the image of the item) as the privileged information based on the existing components, which helps to enhance the model's ability to mine visual information. As such, it allows the original visual features to be fully utilized to achieve the best cold-start recommendation results.

\end{itemize}

\subsection{Case Study}
In this section, we attempt to investigate how the CIERec facilitates the learning process of the multi-modal representation in the embedding space. To this end, we randomly selected five users and items they interacted with to explore how these embeddings varied in different methods. 

As shown in Figure \ref{fig:tsne}, the relevance of users and items is well reflected in the t-SNE space, namely, the more relevant representations are embedded in the more similar positions. We found that naive cross-modal inference leads to a collapse of its training procedure, whereas the embedding representations learned by CIERec show a significant clustering effect, that is, points with the same color tend to form clusters. These observations demonstrate that CIERec is able to effectively facilitate the learning process of cross-modal representations by augmenting content information, so that the representation of users and the items they interact with tend to be close to each other, which may be one of the reasons for CIERec's superior performance.

\section{Conclusion}
This paper proposes a novel cross-modal content inference and feature enrichment recommendation framework, CIERec, which conducts the cross-modal inference from the visual space to the semantic space based on the items' prior knowledge, and combines a multi-modal representation fusion method to trade-off the heterogeneous representation modeling process from the multi-modal information. Experimental results demonstrate that the introduction of cross-modal inferred information is able to improve the items' representation from multiple perspectives, which makes CIERec superior to existing methods in cold-start recommendation.

Future work of this study will focus on two main directions. First, the heterogeneous alignment techniques may help to model the mapping relationships between visual features and cross-modal inferred features at a multi-granularity level, leading to an increased information gain. Second, we will further improve the representational capability of CIERec by filtering noisy information in cross-modal inference with the graph convolutional network.

\section*{Acknowledgment}
This work is supported in part by the National Natural Science Foundation of China (Grant no. 62006141), the Excellent Youth Scholars Program of Shandong Province (Grant no. 2022HWYQ-048), and the TaiShan Scholars Program (Grant no. tsqn202211289).

\balance
\bibliographystyle{IEEEtran}
\bibliography{reference.bib}

\end{document}